# Observable effects of Interplanetary Coronal Mass Ejections on ground level neutron monitor counting rates

J. J. Blanco, E. Catalán, M.A. Hidalgo, J.Medina, O.García and J. Rodríguez-Pacheco

Space Research Group, Physics Department, Alcalá University, Ctra. Madrid-Barcelona km 33.600. E-28871, Alcalá de Henares (Madrid), Spain

**Abstract**

In this work, non-recurrent Forbush decreases (FDs) triggered by the passage of shock-driven interplanetary coronal mass ejections (ICMEs) have been analyzed. Fifty-nine ICMEs have been studied but only the 25% of them were associated to a FD. We find that shock-driving magnetic clouds (MCs) produce deeper FDs than shock-driving ejecta. This fact can be explained regarding to the observed growing trends between decreases in neutron monitor (NM) count rate and MC/ejecta speed and its associated rigidity. MCs are faster and have higher associated rigidities than ejecta. Also the deceleration of ICMEs seems to be a cause in producing FDs as can be inferred from the decreasing trend between NM count rate and deceleration. This probably implies that the interaction between the ICME traveling from the corona to the Earth and the solar wind can play an important role to produce deeper FDs. Finally, we conclude that ejecta without flux rope topology are the less effective in unchaining FDs.

**Introduction**

Ground level neutron monitors (NMs) are able to monitor the galactic cosmic ray (GCR) fluxes arriving to the Earth surface with energies between 0.5 to 20 GeV (Simpson, 2000). The geographical location of a NM determines the minimum energy of GCRs that reaches each station. This is traditionally quantified by the geomagnetic cutoff expressed in GV. Particles with less magnetic rigidity than the NM geomagnetic cutoff cannot reach the monitor. The NM count rate can be strongly affected by solar flares (Firoz et al., 2011), coronal mass ejections (CMEs) (Gopalswamy et al., 2012) and solar wind structures such as interplanetary coronal mass ejections (ICMEs) (Jordan et al., 2011), interplanetary shocks (Cane, Richardson and von Rosenvinge, 1994) and interaction regions (Richardson, Wibberenz and Cane, 1996). While the two formers can produce a significant increase in the NM count rate, known as ground level enhancement (GLE) (Shea and Smart 2012), the other three may induce decreases in NM count rate called Fosbush decrease (FD). These FDs can be divided into recurrent or non recurrent depending on if they are observed along to several solar rotations and are associated with corotating stream interaction regions (Richardson, Wibberenz and Cane, 1996) or if they last for several days and are caused by transient events as interplanetary shocks or ICMEs passages (Cane, 2000 and Belov, 2008). In this work we focus on non-recurrent decreases and we will refer to them as FDs. A Forbush decrease (FD) is observed as a decrease in the cosmic ray intensity and was first reported by Forbush (1937). It is characterized by a fast decrease, as much as 20% in the order of hours, and a slow recovery phase that can last several days. As a first approach, it can be assumed that the decreases in the cosmic ray counts are due to changes in the propagation conditions at the surrounding region where the FD is observed. It can be said that FD is a local phenomenon restricted to small regions when compared with the whole heliosphere. These changes can be related to enhancements in solar wind speed, variation in the magnetic field topology, enhancements in the interplanetary magnetic field magnitude and the presence of magnetic turbulence. ICMEs are large structures (around 0.1 AU) that propagate at high speed (up to 2000 km s$^{-1}$) and produce shocks and magnetic turbulences on the underlying solar wind. Moreover, about one third of ICMEs shows a closed magnetic topology defined by a relatively strong magnetic field and a smooth field rotation which is usually known as magnetic cloud (MC) (Burlaga et al., 1981; Lepping, Jones and Burlaga, 1990). It is generally accepted that an ICME passage can produce decreases in the count rate of NMs (Cane, 2000; Ifedili, 2004; Papaioannou et al., 2010). These decreases are short term events with the decreasing phase lasting for about one to two days and the recovery phase over one week of duration.

During a shock-driving ICME passage, the shock may initiate a decrease in NM counts maintained along the sheath region, *i.e.*, the highly turbulent region between the shock and the ICME. This decrement can be steeper at leading edge of the ICME. This scenario is path dependent. This means that depending on the trajectory of the spacecraft or the Earth through the shock/ICME structure some of these two effects might not be observed (Richardson and Cane, 2011).

The FD shape may vary from one event to other, especially if complex structures converge on the observation point. Jordan et al., (2011) point out that each FD has to be studied separately and that small scale structures, between shock and ICME, can greatly affect the FD shape and question the two-steps FD picture.

Under the question "do all the CMEs have flux rope structure?" a list of 59 shock-driving ICMEs extracted from Gopalswamy et al. (2010) during Solar Cycle 23, was proposed to be analyzed during the Living With a Star Coordinate Data Analysis Workshop hosted in San Diego (2010) and Alcalá (2011). The subset (these 59 events) was selected using a CME source region criterion (E15° ≤ source longitude ≤ W15°). This roughly implies that only CMEs from the central solar disc region were considered. Although the main goal in this workshop was the study of the magnetic structures observed into ICMEs, we investigated the role of these structures on the propagation of cosmic rays, especially during their arrival to Earth.

In this work we analyze the effect of the ICME passage on Oulu NM station count rate with the goal to study which part of an ICME, *i.e.*, shock, MC or flux rope, magnetic field magnitude and induced turbulence plays the most important role in producing observable FDs.

**Data analysis**

The 59 shock-driving ICMEs selected from the list in Gopalswamy et al., 2010 has been studied from January 1997 to September 2006. In 24 of them, clear signatures of MC were found. We considered that an MC has been detected when the solar wind follows the Burlaga criteria (Burlaga et al., 1981; Lepping et al., 1990), *i. e.* low temperature, smooth magnetic field rotation combined with intense magnetic field, and the magnetic field can be fitted with the Hidalgo's model (Hidalgo and Nieves-Chinchilla, 2012). The other 35 events did not show clear evidences of MC but a depression in solar wind proton temperature is observed with low plasma beta. Generally speaking we named them ejecta (Ej). If the magnetic field into the Ej is organized as a flux rope that can be fitted by the Hidalgo's model, then this Ej is cataloged as ejecta plus (Ej+), and ejecta minus (Ej-) in the opposite case. The ICME pool was separated into MC (24), Ej+ (23) and Ej- (12). The details of this classification can be found in Hidalgo, Nieves-Chinchilla and Blanco (2012).

Key parameters with a time resolution of 92 s from *Solar Wind Experiment* (SWE) (Ogilvie et al., 1995), one minute time resolution data from *Magnetic Field Instrument* (FMI) (Lepping et al., 1995) on board *Wind* spacecraft, 64 s time resolution data from the *Solar Wind Electron, Proton, and Alpha Monitor* (SWEPAM) (Mc Comas et al., 1998), and 4 min resolution data from magnetic field experiment (MAG) (Smith et al, 1998) on board ACE spacecraft have been used. Data have been retrieved from CDAWeb wed page. In this work, it is assumed that a FD is observed when the NM count rate decreases more than 3% below the GCR background measured before the shock arrival. Because of its relative low geomagnetic cutoff (0.81 GV), arriving cosmic rays with energies higher than some hundreds of MeV are detected by the Oulu (Finland) NM. Counts of 5 min of time resolution from Oulu have been used (Kananen et al., 1991). This station is located at 65.05ºN, 25.47ºE and at 15 m above sea level. The monitor is made up by 9 NM-64 tubes. The data from this station have been collected from the Neutron Monitor Database (NMDB) (Mavromichalaki et al, 2011) that integrates the readings of many different NM stations located mainly in Europe and Asia. The high count cadence let us perform comparable observations with measurements acquired by spacecraft instruments with similar time resolution to the one used in our analysis of MC magnetic structure. Although 5-min NM data are not the standard approach to study FDs, where hourly averaged measurements are commonly used (e. g. Cane, 2000; Usoskin et al., 2008; Papaioannou et al., 2010; Richardson and Cane, 2011), this high count cadence is required to make a direct comparison between the results given by Hidalgo's model, *i.e.* MC and/or Ej+ existence and limits, and the role of MCs, Ej+ and Ej- on the depth of FDs.

Only fifteen ICMEs from the selected sample of 59, triggered the detection of a FD in Oulu NM. Eight of them were MC and other six Ej+. Only one Ej- was able to induce an observable FD at Oulu. The latter Ej was preceded by a strong interplanetary shock. It is clear that flux ropes (MC or Ej+) within ICMEs play a crucial role in producing FD (94% of FD associated with flux ropes). The decrease percentage of the compiled FD was ranged between 5.2% and 26.1%, those related to MCs being deeper (Table 1). The transit time, *i. e.* the time that a CME takes to arrive to the Earth, has been calculated using the onset times from the LASCO CME list (http://cdaw.gsfc.nasa.gov/CME_list/) and the ICME *in-situ* times using measurements from instruments on board *Wind* and ACE spacecraft. From Table 1, it seems that the shorter the travel time the deeper the FD. This will be discussed in next section. In our list of 15 ICMEs connected to CMEs from the central region of solar disc, the deepest GCR decrease rate was measured in eight events during the ejecta´s passage, in six events during the ICME sheath and one behind the ejecta's passage. Only in three events, 12 April 2001, 29 October 2003 and 10 November 2004, the FDs could have been affected by other structures. The former two during their recovery phases, because of the presence of a subsequent interaction region and a later ICME, respectively, and the third during its main phase due to a previous ICME which reduced the GCR level before the 10 November MC arrival. Five events produced decreases higher than 10%. All of them were ICMEs with MC and in three of them some interaction with previous or subsequent structures might have happened, as it has been explained above.

For every event, the shock strength, the Ej size, its mean speed and mean magnetic field have been computed. As an example the analysis of two events is shown in detail. On 15 May 2005 a shock arrived at *Wind*'s location ($X_{GSE}$=200 $R_E$, near L1) followed by a sheath and three hours later by an ICME with a MC structure. The ICME front is marked by a jump higher than 30 nT and a fast field rotation (less than 4 hours) characterized by an elevated thermal speed. This region coincides with the deepest point in the FD measured by Oulu NM. At the MC nose, *i.e.* when the field begins to rotate due to the MC passage, a magnetic field intensity of 55 nT and a speed of 990 km s$^{-1}$ was observed. Under these conditions, the ICME reached the Earth 23 minutes later. The shock arrival to the Earth was observed in coincidence with a steep decrease in the counts measured at Oulu, triggering a clear FD (Figure 1 bottom panel where 5 min Oulu NM data have been smoothed using 1 h running average). At ICME front arrival, the FD slope changed being clear the second step in this particular FD. In the figure divided into panels is shown from top to bottom, (1) solar wind density, (2) thermal velocity, (3) solar wind speed, (4) magnetic field components in GSE system (red circles Bx, green triangles By and blue squares Bz) over plotted with continuous lines are the Hidalgo's model results, (5) the magnetic field strength and (6) the percentage of the normalized

NM count rate. The decrease was even steeper when the ICME leading edge hit the Earth. The FD deepest point was measured within the fast rotating region before the MC nose arrival and a soft recovery phase started during the MC passage (marked with vertical lines in Figure 1). The FD lasted more than 5 days until the previous neutron monitor count rate was recovered (doy 140, not shown in Figure 1). The MC showed a well organized magnetic flux rope. This is clear when comparing to the over plotted continuous lines which show the Hidalgo's model result. During this FD, the count rate dropped up to a 12% with respect to the GCR background before the shock arrival, following, in our opinion, a two-step FD shape. This event has been analyzed by Dasso et al. (2009) in terms of the magnetospheric response. They argue the presence of two consecutive MCs. The first one in coincidence with the fast rotating region mentioned above and the second one with the MC presented in Figure 1. In our opinion, the temperature is too high to be sure that such rotation may be a small MC.

On 22 January 2004 a shock arrived at *Wind*'s location in (-210, 42, -2) Earth radii in GSE coordinate system, *i.e.* on the night side of the Earth. To avoid possible interactions with the magnetotail, data from the ACE spacecraft has been used to analyze this event. ACE was located at L1. This event is shown in Figure 2. Plots are organized in as than in Figure 1. There is one difference though. In the second panel from the top appears proton temperature instead of thermal speed. As in Figure 1, the Oulu count rate has been smoothed using running average of one hour to get a clearer structure of the FD. The FD started with the interplanetary shock arrival. During the sheath between the shock and the ICME front the NM counts were reduced by a 4%. Six hours later, when the ICME arrived, a change in the FD slope was detected. Two hours later the flux rope nose, confirmed by Hidalgo's model (continuous lines in Figure 2), was observed. The FD minimum and the beginning of the recovery phase happened within the flux rope. The Ej was characterized by a mean magnetic field of about 10 nT with a smooth field rotation that lasted almost one day and a solar wind speed of 600 km s$^{-1}$ in a low solar wind temperature region. As for the FD shape, it showed a two step behavior with a harder slope in coincidence with the ICME leading edge passage. The recovery phase was slower than that seen in Figure 1, lasting up to 10 days. The neutron monitor registered a decrease of 10% of its counts compared to the GCR background on January 21.

**Results**

The FD depth can be influenced by various ICME properties. One of the possible causes of producing FD can be the size of the magnetic structure and the intensity of its magnetic field. Cane (1993) found a clear correlation between the percentage decrease of GCRs and the magnetic field strength in the ICME. The effect of these two elements can be evaluated by the expression ($R = B\ r\ c$) which gives the magnetic rigidity in GV, *B* being the magnetic field intensity, *r* the particle giroradii and *c* the light speed. We assume the value of *B* as the mean value inside the ICME and *r* as the size of the ICME section because the particle giroradii has to be in the order of this size to be affected in its normal movement. In a recent paper, Kubo and Shimazu (2010) analyzed the effect of finite Larmor radius on GCR penetration into flux ropes concluding that it can be relevant at 1 AU. The mean B and the structure size have been computed using Hidalgo's model both for MCs and Ej+. As for the only Ej-, its size was assumed to be equal to the ICME size. The resulting plot of the FD minimum versus the estimated rigidity is presented in Figure 3a. Red circles represent MCs, blue triangles Ej+ and the green square Ej-. The growing trend of GCR count rate percentage with rigidity is clear, ICMEs with MC being more effective than Ej+ and Ej- in producing FDs. This can be understood as because of the larger MC sizes and the more intense magnetic fields imply higher associated rigidity. One of the MCs (the 29 October 2003 event) showed a percentage decrease higher than 25%. Nevertheless, its rigidity was relatively low. This event had a sheath with a magnetic field as high as 50 nT and an MC mean field of only 12 nT. In this event, the role of the sheath seems to be more important than that of the MC in terms of reducing the Oulu NM count rate.

It can be argued that the shocks observed ahead some ICMEs play an important role in the FD depth themselves, but what we observed in Figure 3b is that those shocks associated with MCs are related to deeper FD. The shock strength is defined here as the ratio of the difference between the downward and the upward magnetic field at shock passage. It is important to point out that the shock driven by the Ej- (green square) was the third more intense, but it only caused a modest FD of 7%. The conclusion that can be extracted from Figure 3 is that an MC strengthens the shock effect on the neutron monitor count rate. Red continuous line and blue dashed line are the linear fits to MCs and Ej+ with slopes of 13.6 and 4.7 and Pearson's coefficients (Pc) of 0.76 and 0.66 respectively. The shock triggers the FD but the MC makes it deeper. This result is in agreement with Richardson and Cane (2011) concerning the role that MCs may play in producing FD. Also the observed relationship between Ej rigidity and FDs could support the argument of MCs being closed magnetic structures.

Other important parameter that deserves to be studied focusing on the causes of FDs is the speed of the ICME. There are three different speeds that can be associated with the ICME propagation. The CME emergence speed that is calculated from coronagraph images, the ICME transit speed that can be estimated from the CME onset time and the ICME arrival time at the spacecraft location and the solar wind speed measured within the ICME. A common conclusion inferred from the three speeds is that the faster CME or ICME the deeper the FD (Figure 4). This result agrees with those by Richardson and Cane (2011) who used a pool of more than 300 ICMEs. As they affirm in their paper, the decrease dependence on CME/ICME/MC speed can be explained arguing that in faster-propagating events GCRs have less time to fill up the closed magnetic structure on an MC. On the other hand, the range of values of the three speeds is different. The CME speed is ranged between 300 and 3000 km s$^{-1}$ (Figure 4a), the transit speed between 500 and 2000 km s$^{-1}$ (Figure 4b) and the solar wind speed between 300 and 1300 km s$^{-1}$ (Figure 4c). Again, MCs are, generally speaking, faster than Ej. On the other hand, taking a closer look at Figure 4c, it is clear that Ej velocities are in a narrow range of 500 km s$^{-1}$ (250 to 750 km s$^{-1}$) without a clear linear relationship (blue dashed line) with the associated FD (Pc = 0.33). Nevertheless, the FDs produced by MCs show a good linear correlation (red continuous line) and a clear growing trend with the MC speed at 1 AU (Pc = 0.71).

Non-recurrent FDs are observed by NMs at ground level as local phenomena related to solar wind conditions around Earth, given that most of them can be directly related to the passage of an ICME. No relationship of CME speed, transit speed and FD should be expected other than the dependence between these two velocities on the solar wind speed. Nevertheless, important variations in the speed from the CME onset to the ICME arrival at the Earth are depicted in Figure 4. This can be explained by assuming that an effective interaction between ICMEs and solar wind occurs during the ICME's travel in the interplanetary space (Vrsnak, 2001). In almost all the events a deceleration is observed. This deceleration can be due to an effective kinetic energy exchange between the ICME and the solar wind. This exchange can produce intense shock waves and turbulences ahead (sheath) the ICME and therefore make the ICME able to change the propagation conditions of GCRs with energies from hundreds to thousands of MeV. This is expected for propagating diffusive barriers (Wibberenz et al., 1998). The ICME acceleration can be estimated from the difference between the solar wind speed and the CME speed divided by the travel time. In Figure 5, this acceleration is plotted against the percentage decrease of GCRs displaying a clear negative slope. Those ICMEs that are more intensively decelerated produce deeper FDs. Only two of fifteen ICMEs show positive acceleration. Although acceleration could produce an effective interaction with solar wind, two events provide a little statistical evidence to affirm that a change in the acceleration slope appears. Moreover, this cannot be considered as a conclusive result because of the uncertainty in CMEs speed estimations. MCs and Ej+ events have similar slopes (red continuous and blue dashed lines respectively) of 0.04 and 0.035 but different Pcs, -0.89 for MCs and -0.53 for Ej+. MCs are more efficiently decelerated. This may be due to their larger size, more intense magnetic field and higher speed. Moreover, the FD depth is better correlated with MC acceleration than with any other physical quantities considered in this work (rigidity, shock strength, speed). The deceleration/acceleration of ICMEs plays a very important role in the development of FDs. Intense accelerations imply strong interaction between ICME and solar wind. This interaction drives stronger shocks and makes the solar wind more turbulent. These two features affect greatly the propagation of cosmic rays in the range of the detectable energies by the neutron monitors.

**Conclusions**

The role of CMEs originating from near the center of the solar disc and their associated ICME on FDs detected by the Oulu NM has been analyzed. Cosmic rays with energies higher than a few hundreds of MeV are the main component of the energetic particle population detected by this NM. A pool of 59 shock-driving ICMEs have been classified into three groups, MC (24), ejecta with flux rope (23) and ejecta without apparent flux rope structure (12). Only around 25% of them were able to produce decreases in the NM count rate higher than 3%, eight MC, six Ej+ and one Ej-. This result seems to show that an isolated shock is rarely able to produce FD. Moreover, similar shocks may induce stronger FDs if they are driven by an MC or an Ej+. Therefore a closed magnetic structure such as MC or flux rope strengthens the effect of shocks on FDs. Richardson and Cane (2011) reached the same conclusion.

The rigidity associated with MCs and Ej affects the CGR propagation into ICMEs. This rigidity has been compared with the GCR decreases concluding that higher rigidities are related to deeper FD. The higher rigidities correspond to MCs because they are larger and their magnetic fields are more intense than those of the Ej events.

The shock strength and its relationship with FD have been also analyzed. Stronger shocks produce higher decreases in the GCR count rate but when considering similar shocks, those driven by MC are more effective (almost three times more effective) in shielding the Earth from the arriving GCRs. This can be explained assuming that MCs interact more strongly with the underlying solar wind than Ej driving turbulences into the sheath region and therefore, affecting in a more efficient way, the GCRs propagation into the ICME.

Other analyzed aspect is the role of the ICME speed on GCR count rates. The observations show that faster structures (MC or Ej) are more efficient to produce FDs, and at least in the sample analyzed, MCs are faster than Ej events. Moreover, FDs associated with ejecta show an increasing trend with CME speed and transit speed but not so clear with their measured speed at 1 AU (Pc = 0.33). As for MCs these three velocities show similar increasing trends with the FD depth and a good correlation between MC speed and FD depth (Pc = 0.71). This result is also in agreement with the conclusion by Richardson and Cane (2011).

Finally, we have observed that the deceleration/acceleration of ICME between the Sun and the Earth can play an important role in the development of FDs. Higher decelerations induce deeper FDs. This can be explained in terms of effective energy exchange between the ICME and solar wind. This interaction can lead to the formation of a stronger shock ahead the ICME. MCs decelerate stronger. Closed magnetic structures as MCs with stronger magnetic field and larger size than those observed in Ej events seem to be more effective to interact with solar wind. Moreover, we find the best correlation between deeper FDs and the MC acceleration. The linear correlation gives us a Pc equal to -0.89. This value implies that the interaction between MC and solar wind is very important in the shielding effect that an ICME has over GCRs.

Richardson and Cane (2011) propose that MCs are effective in excluding GCRs because they are closed magnetic structures. Our results support this conclusion but also the importance of MC/solar wind interaction on GCR decreases as can be inferred from the clear relationship between MC acceleration and GCR count rates.

Hidalgo, Nieves-Chinchilla and Blanco (2012) have found that most of the ejecta from the initial list of 59 shock-driving ICMEs showed axes close to the Sun-Earth line. This implies that the passage of the spacecraft through the corresponding ejecta event was probably by its flank and it can support the idea of MCs and Ej events are different parts of the same flux rope. According to this picture and the results showed in this work we conclude that the effect of shock-driving ICMEs on GCR count rates may also depend on which region of the flux rope is observed.

In conclusion, shock-driving MCs produce deeper FDs than Ej+ and Ej- events because the MCs have higher rigidity, higher speed and higher deceleration and they interact with solar wind more effectively.

## Acknowledgements

We acknowledge the NMDB database (www.nmdb.eu), founded under the European Union's FP7 programme (contract no. 213007) for providing data, especially to Oulu neutron monitor station and the Sodankyla Geophysical Observatory of the University of Oulu for the operation of the monitor, also to the MFI and SWE instruments on board Wind and Mag and SWEPAM on board ACE and Coordinated Data Analysis Web (CDAWeb) for the use of data. This work has been supported under the grants: JCCM PPII10-0150-6529 and AYA2011-29727-C02-01. Authors wish specially thank to Parque Científico y Tecnológico de Guadalajara (Guadalab) team. This is an unpublished personal copy. The final version of this paper has been published in Solar Physics. The final publication is available at link.springer.com (DOI: 10.1007/s11207-013-0256-1).

**Figure captions**

Figure 1. Example of an ICME with an MC. Data from *Wind* spacecraft and Oulu neutron monitor has been used.

Figure 2. Example of an ICME with Ej+. Data from ACE spacecraft and Oulu neutron monitor has been used.

Figure 3. FD dependence on rigidity (3a) and shock strength (3b).

Figure 4. FD dependence on of CME speed (4a), transit speed (4b) and solar wind speed measured during an ICME passage (4c).

Figure 5. FD dependence on ICME acceleration. This acceleration has been estimated by means of the expression: $(V_{SW}-V_{CME})$ x ICME travel time.

Table 1: ICME associated with FD. Year, time interval between the MC nose and its rear as estimated by Hidalgo's model. CME transit time, magnetic rigidity cut off associated with the flux rope, FD and FD location. The asterisk in MC* means a complex event where two consecutive MCs were observed but the FD is not resolved into two separate events.

Table 1: ICME associated with FD. Year, time interval between the MC nose and its rear as estimated by Hidalgo's model. CME transit time, magnetic rigidity cut off associated with the flux rope, FD and FD location. The asterisk in MC* means a complex event where two consecutive MCs were observed but the FD is not resolved into two separate events.

| Year | ICME interval (doy) | Type | Travel time (day) | Rigidity (GV) | FD (%) | FD location |
|---|---|---|---|---|---|---|
| 1998 | 124.442→125.234 | Ej- | 1.95 | 167 | 7.5 | Ejecta |
| 1999 | 178.942→179.108 | Ej+ | 3.38 | 27 | 6.24 | Sheath |
| 2000 | 197.911→198.298 | MC | 1.46 | 419 | 16.00 | Ejecta |
| 2000 | 261.221→262.599 | MC | 1.77 | 283 | 8.86 | Ejecta |
| 2000 | 311.964→312.74 | MC | 3.2 | 215 | 6.96 | Ejecta |
| 2000 | 332.458→333.131 | Ej+ | 3.23 | 80 | 9.01 | behind Ej+ |
| 2001 | 102.367→103.279 | MC* | 2.14 | 139 | 12.63 | Ejecta |
| 2001 | 118.892→119.662 | MC | 2.37 | 106 | 8.17 | Sheath |
| 2001 | 285.205→285.360 | Ej+ | 2.73 | 24 | 7.67 | Sheath |
| 2003 | 302.554→303.151 | MC* | 1.07 | 162 | 26.13 | Ejecta |
| 2004 | 22.558→23.282 | Ej+ | 2.55 | 142 | 10.09 | Sheath |
| 2004 | 315.195→315.705 | MC* | 4.11 | 188 | 12.42 | Sheath |
| 2005 | 135.464→136.004 | MC | 1.75 | 354 | 11.89 | Sheath |
| 2005 | 149.495→149.638 | Ej+ | 2.6 | 27 | 6.10 | Ejecta |
| 2006 | 232.630→233.625 | Ej+ | 3.94 | 80 | 5.24 | Ejecta |

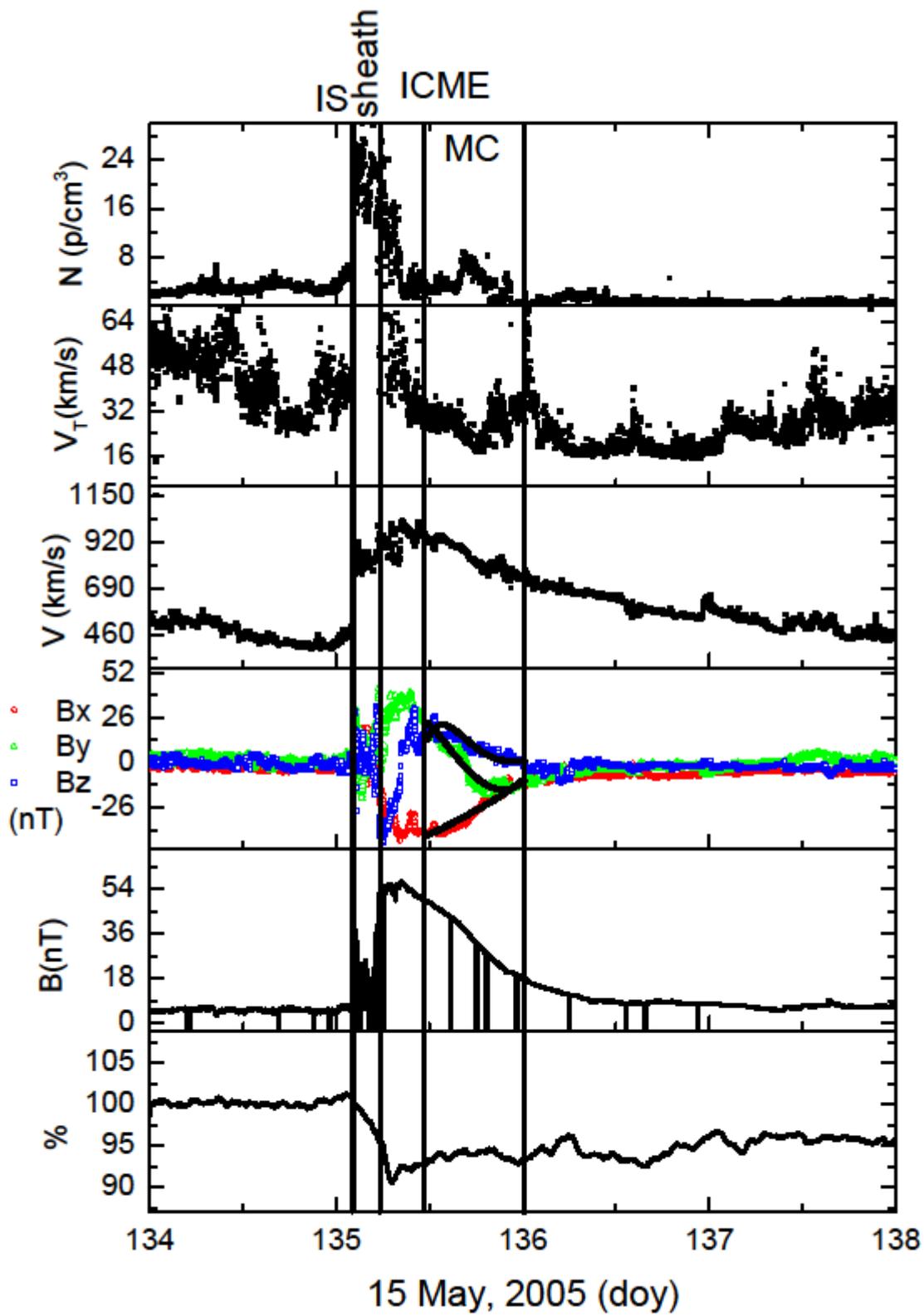

Figure 1:

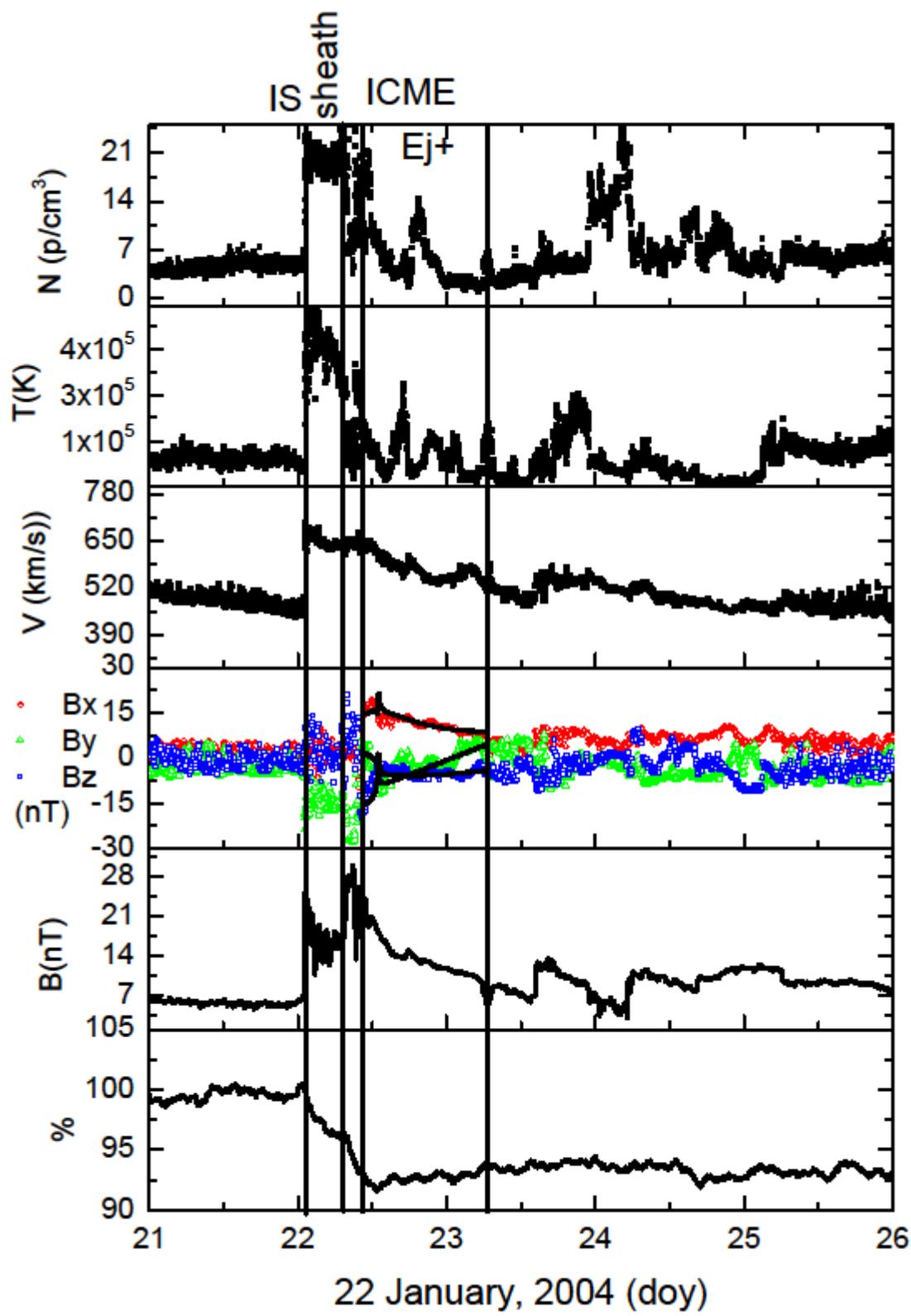

Figure 2:

Figure 3:

Figure 4:

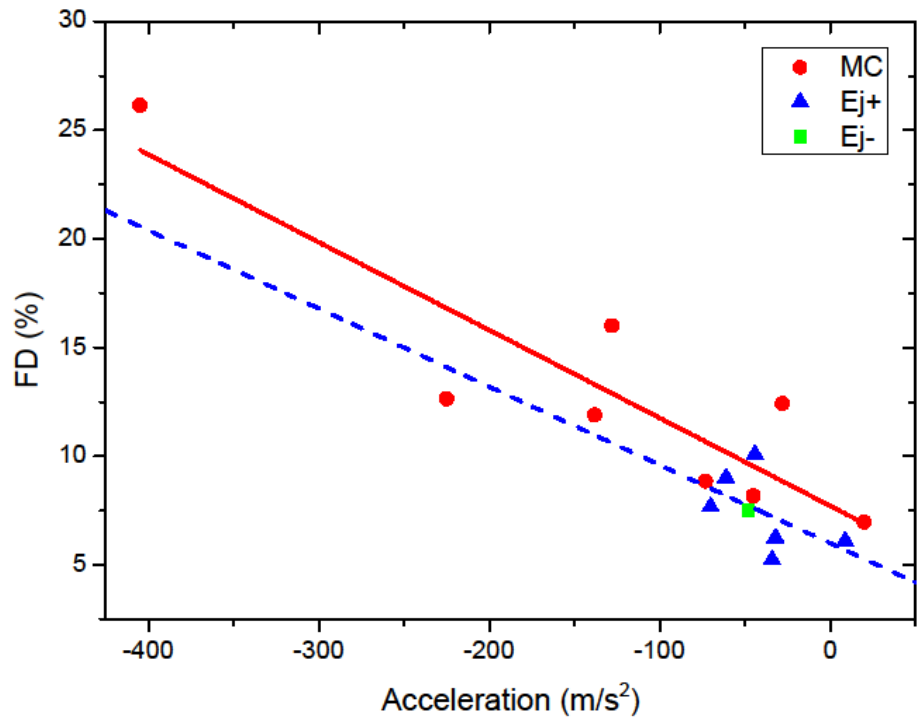

Figure 5: